\documentstyle[epsfig]{aipproc}

\def\'#1{\ifx#1i{\accent"13\i}\else{\accent"13#1}\fi}
\def\etal{{\it et al.}}

\def\rhot{\rho_{\rm t}}

\def\VS{V\'azquez-Semadeni}

\begin{document}
\title{Virial Balance in Turbulent MHD Two Dimensional
Numerical Simulations of the ISM
\thanks{This work has received
partial finantial support from grants UNAM/CRAY grant {\tt SC-002395},
UNAM-DGAPA {\tt IN105295}, UNAM-PADEP grant {\tt 003319} and scolarship from
UNAM-DGAPA.}
}

\vskip -1.5cm

\author{Javier Ballesteros-Paredes and Enrique
V\'azquez-Semadeni.}
\vskip -1cm
\address{
Instituto de Astronom\'\i a, UNAM. Apdo.\ Postal 70-264, M\'exico, D.F. 04510.
{\tt javier@astroscu.unam.mx}, {\tt enro@astroscu.unam.mx}}

\lefthead{Ballesteros-Paredes \&\ V\'azquez-Semadeni}
\righthead{Virial Balance}
\maketitle
 \vskip -1cm
\begin{abstract}
We present results from a virial analysis of fully nonlinear two-dimensional
(2D) simulations of the ISM. We discuss the Eulerian Virial Theorem
in 2D, and describe preliminary results on the virial budget of clouds
in the simulations. The clouds are far from a static equilibrium, and the
Virial Theorem is dominated by the time-derivative terms, indicating the
importance of flux through the cloud boundaries and mass redistributions.
A trend towards greater importance of the gravitational term at larger scales
is observed, although a few small clouds are strongly self-gravitating.
The magnetic and kinetic terms scale linearly with each other.
\end{abstract}

\section{Introduction}

V\'azquez-Semadeni \etal\ (1995, hereafter Paper~I) and Passot
\etal\ (1995, hereafter Paper~II) have produced a numerical model of the
interstellar medium (ISM) including enough physical agents as to render it
feasible to perform statistical studies of the clouds formed in the
simulations. The simulations include self-gravity, magnetic fields,
parameterized cooling and diffuse heating, the Coriolis force, large-scale
shear, and localized stellar energy input. In the present work, we discuss
the Virial Theorem (VT) as it applies to the simulations, and present
preliminary statistical results from a two-dimensional (2D) simulation with
a resolution of $800 \times 800$ grid points, performed specifically for
this analysis. In \S~II we discuss the VT, applying the formalism developed
by McKee \& \ Zweibel (1992) to the 2D case. In \S~III we describe the
cloud-identifying algorithm and show preliminary statistical results, and
in \S~IV we present some remarks and future work.

\vskip -1cm

\section{VIRIAL THEOREM IN 2D}

\vskip -0.3cm

The VT is obtained by dotting the momentum equation (eq. (1b) in Paper~I)
with the position vector {\bf x} and integrating over volume.
McKee \& \ Zweibel (1992) have discussed an Eulerian
form of the VT, which is most appropriate for our simulations,
since they are performed with an Eulerian code.
Because the simulations are 2D (in order to reach a sufficiently large
resolution), we must consider the VT in 2D as well. It reads:


\begin{equation}
{1\over 2} {d^2I\over dt^2} =
2 \biggl(\tau_{\rm kin} + \tau_{\rm int} \biggr) + M - W - E_{\rm cor} -
{1\over 2} {d\Phi \over dt}
\label{virial}
\end{equation}
where
$\tau_{\rm kin} = 1/2 (\int \rho u^2 dV - \oint_X x_i \rho u_i u_j \hat
n_j dS )$
is the kinetic term,
$\tau_{\rm int} = \int P dV - 1/2 \oint_S P x_i \hat n_i dS$
is the thermal term,
$M = {1/8\pi} \oint_S x_i  T_{ij} \hat n_j  dS$
is the magnetic term,
$W = \int x_i \rho \ g_i dV$
is the gravitational term,
$E_{\rm cor} = 2 \int x_i ({\bf \Omega} \times {\bf u})_i  dV$
is the Coriolis term,
$\Phi = \oint_S \rho u_i r^2 \hat n_i dS$
is the flux of moment of inertia through the surface $S$, and
$\rho$, ${\bf u}$, $P$ and $g_i$ are the density, velocity, thermal pressure
and self-gravitational acceleration, respectively. Because
of two-dimensionality, we must replace volumes by areas and surfaces by
contours in (\ref{virial}). However, we retain the above notation for
generality. Since in 2D $\nabla \cdot {\bf x} = 2$, in equation (\ref{virial})
we note  the three following interesting points: {\it a)} Although
magnetic fields are present in the surface term
$
M = \int_S x_i \ T_{ij} \ \hat n_j \ dS,
$
(where the Maxwell stress tensor is defined as
$
T_{ij} \equiv 1/4\pi [ B_i B_j - {1/2} B^2 \delta_{i,j} ]
$), the ``classical" magnetic energy term
$
E_{\rm mag} = {1/8\pi} \int B^2 dV
$
does not enter the virial equation, so it does not provide support against
gravity.
{\it b)} The internal energy
$
E_{\rm int} \equiv \int P dV
$
does not contain the 3/2 factor as in 3D. Nevertheless,
in 2D this term still coincides with the total internal
energy, because there are only two translational degrees of freedom.
{\it c)} Additionally, it can be shown that the gravitational term
$\int x_i \ \rho \ g_i \ dV$ does not coincide with the gravitational energy
$E_{\rm grav} = {1/2}\int \rho \  \phi \ dV$ as it does in 3D for isolated
clouds. Essentially, this is due to the slower distance dependence of the
gravitational potential in 2D.

\vskip -0.5cm

\section{PRELIMINARY STATISTICS}

\vskip -0.3cm

\begin{figure}
\def\epsfsize#1#2{1.03\hsize}
\centerline{\epsffile{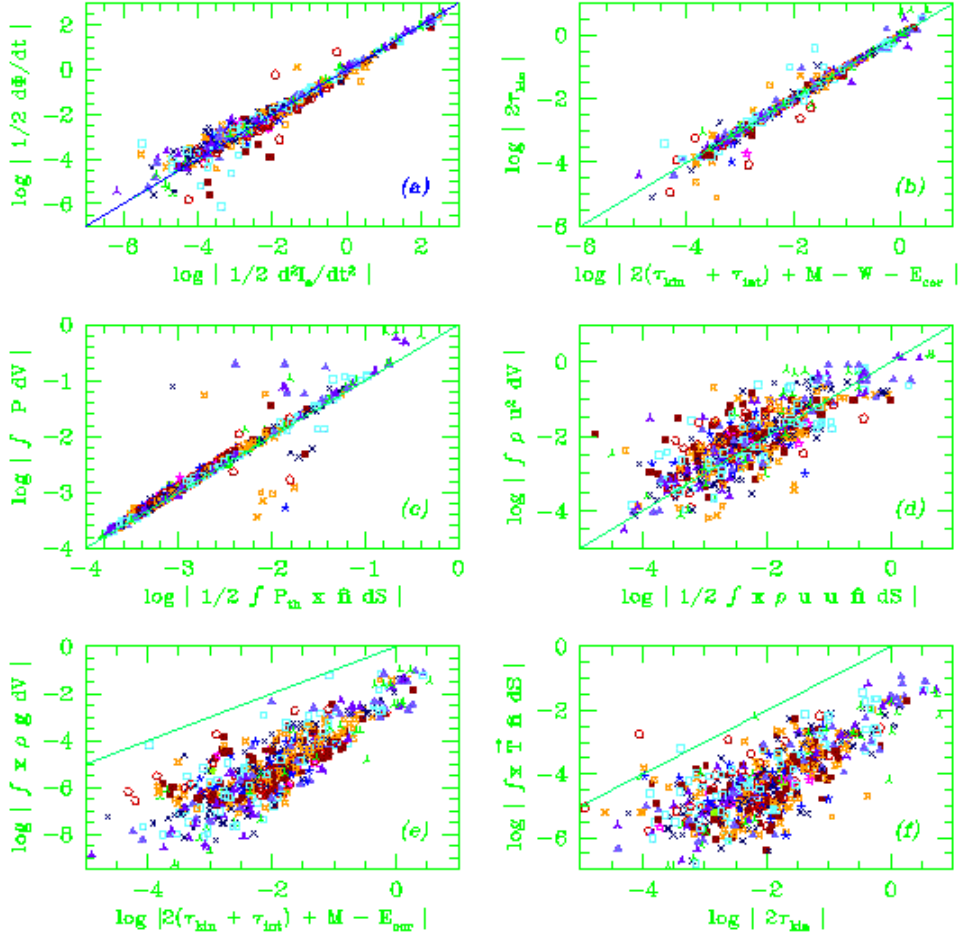}}
\vskip -2cm
\caption{In all panels, the solid line is the identity. {(a)}
$1/2 || d\Phi /dt ||$ vs. $1/2 || d^2I/dt^2 ||$. Their near equality
shows that the
term $1/2 d\Phi /dt$ dominates the virial sum, indicating the importance of
the variability of the mass flux through the clouds' borders for the total
virial balance. {(b)} Volume-plus-surface kinetic terms vs.\ the virial
sum neglecting  the $1/2 d\Phi /dt$ term.
The near equality of both terms indicates the dominance
of the kinetic terms over the remaining ones. This effect may be due to
cloud bulk motion and should be eliminated by using an
instantaneously-at-rest frame of reference for each cloud. {(c)} Volume
vs.\ surface terms for internal energy
(pressure) and (d) kinetic energy. The surface terms are seen to be
comparable to the volume terms in general. The few points with large scatter
in {(c)} are likely to correspond to regions of anomalous pressures due
to recent star formation. {(e)}
The gravitational term $W$ vs. the sum of the remaining virial
terms. A trend towards greater importance at larger scales is seen. However, a
few points at near balance with gravity are seen at all scales.
{(f)} Magnetic term $M$ vs.\ the sum of the kinetic terms. An almost
linear relation is observed. This is consistent with equipartition between
kinetic and magnetic modes, if an offset is present, again
due to the fact that clouds may
have bulk velocities with respect to the integration volume.
}
\label{dominant-terms}
\end{figure}

In order to calculate the terms in equation (\ref{virial}), we have performed
a 2D simulation similar to
the one called ``Run 28'' in Paper~II, but with a resolution of $800 \times
800$ grid points. In this run we analyze
the data shortly after turning off star formation, in order to allow for the
largest possible density gradients (see \VS, Ballesteros-Paredes \&\
Rodr\'iguez 1997, hereafter Paper~III) while still retaining the structure
induced by the stellar energy injection. We have developed a numerical
algorithm
to identify clouds and evaluate within them the various terms entering the VT,
as well as their velocity dispersion and mean density. We define a cloud
as a connected set of pixels whose densities are larger
than an arbitrary threshold $\rhot$.
Previous calculations (Paper~III) have shown that the
simulations exhibit similar scaling properties as those observed in real
interstellar clouds (Larson 1981), except for the density-size scaling
relation, supporting the possibility that it may be the result of an
observational effect (see also Larson 1981, Kegel 1989, Scalo 1990). With
this motivation, we have now performed evaluations of the various terms in
the VT. We have the following preliminary results: {1.-} Both the second
derivative of the moment of inertia and the last term in the equation
(\ref{virial}) are dominant in the overall virial balance (fig.
\ref{dominant-terms}{a}). {2.-}Comparing the remaining terms, the turbulent
terms are seen to dominate (fig. \ref{dominant-terms}{b}). {3.-}The surface
terms (which are often neglected under the assumption of vanishing fields
outside the clouds) are in general of magnitude comparable to that of the
volumetric ones (figs. \ref{dominant-terms}{c} and {d}). {4.-}The
gravitational term is most important at large scales (fig.
\ref{dominant-terms}{e}). However, there are a few small (low energy content)
clouds which have large values of the gravitational term. These may be the
best candidates for collapse and star formation. Their scarcity appears
consistent with the low efficiency of star formation. {5.-}The magnetic term
and the sum of the kinetic terms are proportional to each other (fig.
\ref{dominant-terms}{f}). This suggests there is equipartition between
kinetic and magnetic modes, except for a constant factor, which may
be due to the fact that clouds have bulk velocities with respect to the
integration volume.

\vskip -2cm

\section{FINAL REMARKS}

\vskip -0.3cm

The dominance of the time-derivative and kinetic terms indicates the
importance of flow through the volume boundaries, contrary to the cases
considered by McKee \&\ Zweibel (1992). In order to minimize this effect,
it appears necessary to consider Eulerian volumes instantaneously at rest
with respect to the center of mass of the clouds. However, preliminary
attempts suggest that the flow through the boundaries cannot be eliminated
completely, since the clouds are extremely amorphous and change shape rapidly.
This work will be reported in a future paper (Ballesteros-Paredes \&\ \VS\
1997, in preparation).

\end{document}